\documentclass{sig-alternate-05-2015}
\usepackage{wrapfig}
\usepackage{booktabs}
\usepackage{enumitem}
\usepackage{color}
\usepackage[square,numbers,sort&compress]{natbib}
\usepackage{longtable}
\usepackage{subfig}

\setlist[description]{leftmargin=*}
\setlist[enumerate]{leftmargin=*}
\begin{document}

\setcopyright{acmcopyright}

\doi{http://dx.doi.org/xx.xxxx/xxxxxxx.xxxxxxx}

\isbn{978-1-4503-4486-9/17/04}


\acmPrice{\$15.00}

%
 \conferenceinfo{HIP3ES'2017}{, January 25, 2017, Stockholm, Sweden}
\CopyrightYear{2017} 

\title{Automatic SDF-based Code Generation from Simulink Models for Embedded Software 
Development}

\numberofauthors{2} 
%
\author{
\alignauthor
Maher Fakih\\
\affaddr{OFFIS Institute for Information Technology}\\
\affaddr{Oldenburg, Germany}\\
\email{maher.fakih@offis.de}
\alignauthor
Sebastian Warsitz\\
\affaddr{Carl von Ossietzky Universität Oldenburg}\\
\affaddr{Oldenburg, Germany}\\
\email{sebastian.warsitz@uni-oldenburg.de}
}

\date{25 January 2017}

\maketitle
\sloppy

\begin{abstract}
\footnotesize{
\noindent
Matlab/Simulink is a wide-spread tool for model-based design of embedded systems. 
Supporting hierarchy, domain specific building blocks, functional simulation and automatic 
code-generation, makes it well-suited for the design of control and signal processing 
systems. 
In this work, we propose an automated 
translation methodology for a subset of Simulink models to Synchronous 
dataflow Graphs (SDFGs) including the automatic code-generation of SDF-compatible embedded 
code. 
A translation of Simulink models to SDFGs, is very suitable due to Simulink actor-oriented 
modeling nature, allowing the application of several optimization techniques from the 
SDFG domain. Because of their well-defined semantics, SDFGs can be analyzed at compiling 
phase to obtain deadlock-free and memory-efficient schedules. In addition, 
several real-time analysis methods exist which allow throughput-optimal mappings of 
SDFGs to Multiprocessor on Chip (MPSoC) while guaranteeing upper-bounded latencies.  
The correctness of our translation is justified by integrating the SDF generated code as a 
software-in-the-loop (SIL) and comparing its results with the results of the 
model-in-the-loop (MIL) simulation of reference Simulink models.
The translation is demonstrated with the help of two case studies: a Transmission 
Controller Unit (TCU) and an Automatic Climate Control.
}
\end{abstract}

%


%
%

\section{Introduction}
Model-based Design (MBD) of embedded systems is nowadays a standard,
easy and efficient way for capturing and verifying embedded software functional 
requirements. 
The main idea is to move away from manual coding, and
with the help of mathematical models create executable specifications using a 
certain modeling framework.
These frameworks typically provide automatic code generators which generate consistent 
imperative code ready to be deployed in real environments. 
Matlab/Simulink \cite{mat1} is one of the most wide-spread tools for model-based design of 
embedded systems which combines above features in a single framework. 
Simulink utilizes block-diagram to represent system models at the algorithmic level.
For instance, in case of a control system, the model consists of the controller algorithm 
block which controls the environment block (or the process to be controlled typically 
modeled as a set differential equations).

A translation of Simulink models to Synchronous Dataflow Graphs (SDFGs)  
\cite{LeeM87} which are, opposed to Simulink, formally based, is beneficial. Such 
a translation would pave the way towards the application of several optimization and 
formal verification techniques well-established for the SDFG domain.
For e.g. in a recent work in \cite{fakihJSA2015} the formal 
real-time verification (based on model-checking) of SDF applications running on 
Multiple-Processor-System-On-Chip (MPSoCs) with shared communication resources was 
shown to be more viable than the real-time (RT) verification of generic tasks.
Also for SDFGs deadlocks and bounded buffer properties are decidable 
\cite{LeeM87}. In addition with the help of mathematical methods
easy-to-analyze compile-time schedules can be constructed for SDFGs. 
Furthermore, memory-efficient code optimization are available 
\cite{bhattacharyya_clustering, bhattacharyya_synthesis_1999} to 
enable efficient implementations of embedded systems.

In this paper, we present a translation procedure of a defined subset of 
Simulink models to SDFGs based on the work in \cite{simulink2sdfWarsitz2016}. We extend 
the approach in \cite{simulink2sdfWarsitz2016} by enabling the translation of Simulink 
models with multirates features to SDFGs. In addition,  we integrate the translation 
procedure within Matlab/Simulink and utilize the automatic code-generation feature to 
generate SDF-based code from Simulink models.  Moreover, we enable an 
automatic setup of a verification flow which allows a Software-In-the-Loop (SIL) simulation 
showing the functional equivalence of the generated code to the reference model.
  
The paper is structured as follows. We will first recap the basic concepts of synchronous 
dataflow graphs and Simulink models identifying their main differences. Afterwards, 
we discuss the related work in Sect.~\ref{sec:RelatedWorkDF} mainly addressing
translation approaches of Simulink models to SDFGs. Next we elaborate on our translation 
procedure in Sect.~\ref{subsec:Simulink2sdf}, starting with description of the set of 
constraints on the Simulink model enabling the translation. In addition, we discuss the 
code-generation and SIL verification features. Sect. \ref{Evaluation} demonstrate the 
viability of our translation approach with the help of a Transmission Controller Unit 
(TCU) case study. Finally, we conclude our work and give an outlook on open issues
and future work.
\section{Background}
\subsection{Synchronous Dataflow Graphs}
\label{sec:bg:sdfgs}

\begin{figure}[tb]
	\begin{center}
		\includegraphics[width=0.35\textwidth]{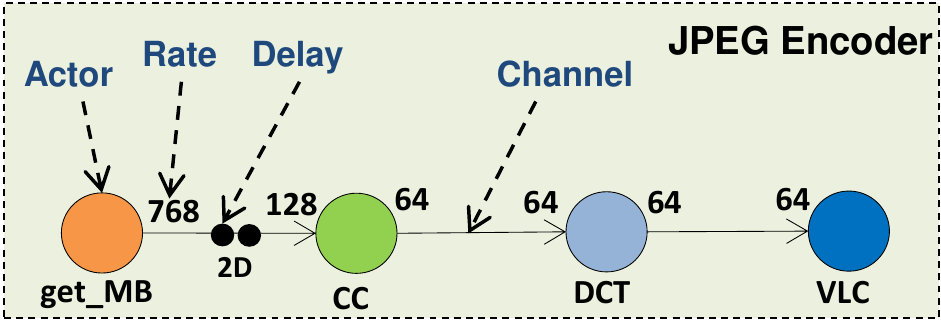}
	\end{center}
	\caption{SDFG of a \textit{JPEG Encoder}}
	\label{fig:sdfg}
\end{figure}

A \textit{synchronous (or static) data-flow  graph (SDFG) }\cite{LeeM87} 
is a directed graph (see Fig.\ref{fig:sdfg}) which, similar to  general data-flow 
graphs (DFGs),
consists mainly of nodes (called \textit{actors}) modeling atomic 
functions/computations 
and arcs
modeling the data flow (called \textit{channels}). In difference to DFGs, SDFGs
consume/produce a static number of data samples (\textit{tokens}) each time an 
actor
executes (\textit{fires}).
An SDFG suits well for modeling multi-rate streaming applications and DSP 
algorithms and
also allows static scheduling and easy parallelization. 
A port \textit{rate} denotes the number of tokens produced or consumed in every 
activation 
of an actor. 
The data flow across a channel (which represents a FIFO buffer) is done according 
to a
First-In-First-Out (FIFO) fashion.
Channels could also store initial tokens (called delays indicated by bullets in 
the edges) 
in their initial state which help resolving cyclic
dependencies (see \cite{LeeM87}).

Despite the analyzability advantage of SDFGs, yet this comes at the cost of their 
expressiveness.
One of the main limitations of SDF Model of Computation (MoC) is that dynamism cannot be handled for e.g.
in the case where depending on the current scenario the application rates changes 
(c.f. \cite{SDFImplementation2013}).
Another limitation (c.f. \cite{LeeM87}) of the SDF MoC is
that conditional control flow is only allowed within an actor functionality but
not among the actors. However, emulating control flow within the SDFG is
possible even though not always efficient (c.f. \cite{SDFImplementation2013}). 
Due to above limitations, for e.g. stopping and restarting an SDFG is not possible 
since an SDFG can have only
two states either running or waiting for input. In addition, reconfiguration of
an SDFG to be able to (de)activate different parts depending on specific modes
is not possible. Moreover, different rates depending on run-time conditions
are not supported. Also modeling exceptions which might require deactivating
some parts of the graph is not possible.
An additional issue is that the SDF model does not reflect the real-time nature of 
the
connections to the real-time environment.

\subsection{Simulink}
\label{sec:bg:Simulink}
Simulink is a framework for modeling of \textit{dynamic systems} and simulating
them in virtual time.
Modeling of such systems is carried out graphically through a graphical editor
consisting mainly of blocks and arrows (\textit{connections}) between them 
representing
signals.
Each block has its input, output and optionally state variables. The relationship 
of the
inputs with the old state variables and the outputs update is realized through
mathematical functions.
One of the powerful features of Simulink is the ability to combine multiple 
simulation
domains (continuous and discrete). This is very useful for embedded systems, where 
in
general the controller has discrete model and the environment  often needs to be 
modeled
as a continuous one.

Simulink also supports a state-based MoC the \textit{Stateflow} \cite{stateflow}  
which is
widely used to model discrete controllers.
Simulink allows a fast \textit{Model-in-the-Loop (MIL)} verification, where the
functional model (of the controller for example) is simulated and results are
documented to be compared with further refinements. In addition, a 
\textit{Software-in-the-Loop (SIL)} verification
is also possible in which the controller model is replaced by the generated code 
from
the \textit{Embedded Coder} \cite{Simulink-coder}  (usually embedded in a 
S-function) and the behavior of the code is
compared with the reference data achieved from MIL (described above).

In \cite{Simulinksdf2} a method was presented to automatically transform SDFGs 
into SBDs (Synchronous Block Diagrams),
such that the semantics of SDF are preserved, and it was proven that Simulink can 
be used
to capture and simulate SDF models. Also authors in \cite{gajski} support this 
fact that dataflow models
fit well to concepts of block diagrams and are used by Simulink. In general, the MoC of Simulink is much 
more expressive than that of the SDF having the advantage of being able to relax 
all limitations of the SDF MoC but at the 
cost of its analyzability.



\section{Related Work}
\label{sec:RelatedWorkDF}
 In the last decade,  several research  \cite{caspi_Simulink_2003, 
miller_formal_2005,
zhang_bridging_2013, buker_automated_2013} have been conducted to enable a 
translation of
Simulink models to other formal models for the purpose of formal analysis. In the
following, we merely discuss  previous work enabling the translation
of Simulink models to SDFGs.

In \cite{GitHubSimulink2SDF}, only the source code of a so-called
\textit{Simulink2SDF} tool was published which enables a very simple translation 
of Simulink models to SDFGs.  In this work all Simulink blocks, without
any distinction, were translated to data-flow actors and similarly connections 
were translated in data-flow channels, the fact which makes the translation  incomplete as 
we will see in Sec.~\ref{subsec:Simulink2sdf}. In addition, our approach allows the 
generation of executable SDF-code which is not possible in this approach.

In \cite{Simulink2sdfBsc_2011}  a translation of Simulink models to
homogeneous SDFGs (HSDFGs) was pursued with the objective of  analyzing 
concurrency.
HSDFGs are SDFGs with the restriction that the number of consumed and produced
tokens of each actor must be equal to 1 \cite {lee_synchronous_1987}. The 
translation has been done for a fixed number of functional blocks but important 
attributes, such as the data type of a connection between blocks, have not been taken into 
consideration by the the translation.

In \cite{Sim2Modal2Ptolmy2SDF2011} it was shown how a case study of a vehicle climate 
control modeled in Simulink is imported to a tool (MoDAL) supporting SDF MoC. MoDAL, in 
turn, exports the model in a format which can be imported by the Ptolemy tool 
\cite{Ptolemy2011}. Ptolemy is then used to generate code from the SDF model. 
In \cite{Sim2Modal2Ptolmy2SDF2011}, only the use-case model have been translated to an SDFG 
without general defining a translation concept applicable at least to a subset of Simulink models.

In \cite{bostrom_contract-based_2015}  a translation from Simulink models to
SDFGs was described. The aim of this work was to apply a methodology for
functional verification of Simulink models based on \textit{Contracts}.
\textit{Contracts} define pre- and post conditions to be fulfilled for programs or
program fragments. 
In \cite{rtasKlikpoKM16}, the ability of SDFGs to model multi-periodic Simulink systems was 
formally proved. There, in addition to systems with harmonic periods, also non-harmonic 
periods are supported (unlike our work and that of \cite{bostrom_contract-based_2015} 
where only harmonic periods are supported).
However,  authors in above work, give no clear classification of critical Simulink 
functional blocks (e.g. the \textit{switch} block with dynamic rates see 
Sec.~\ref{subsec:Simulink2sdf}) 
which cannot be supported in the translation. In addition, \textit {Triggered-/Enabled} 
subsystems and other important attributes such as the data type of a connection are not 
supported. Furthermore, SDF-based code-generation was not considered.

Unlike the above work,  we present a general translation concept based on a 
classification of blocks and connections in Simulink models. Our approach enables the 
translation of critical blocks (such as \textit{Enabled/Triggered} subsystems)
including the enrichment of the translated SDFG with important attributes such as 
the data types of tokens, tokens' size and sampling rates of actors (in case of 
multi-rate models).
This enables a seamless code generation of the model into SDF-based embedded 
software ready to be deployed on target architecture. We also provide an automation of the 
process of SDF-based code generation together with the SIL verification to prove the 
soundness of the translation.


\section{Simulink to SDFG Translation}
\label{subsec:Simulink2sdf}

As already stated (see Sec.~\ref{sec:bg:Simulink}), Simulink MoC is much more 
expressive
than the SDFG MoC. Unlike SDFGs, Simulink supports following additional features:
\begin{description}
\item[U1] \textbf{Hierarchy} (e.g. \textit{subsystem} blocks): While in
Simulink multiple functional blocks  can be grouped into a subsystem,  in
SDFGs each actor is atomic and therefore no hierarchy is supported. \label{U1}
\item[U2]  \textbf{Control-flow logic/Conditional} (for e.g. \textit{switch 
block}
or \textit{triggered subsystem} see \cite{mat1}):  In Simulink control flow is 
supported 
on the block level. This means that  depending on the value of a control signal at 
a block,
different data rates could be output by the block. In contrary, in SDFGs data 
rates
at input and output ports of an actor are fixed and control structures are only
allowed within the functional code of an actor and can't be represented in an 
SDFG.
\label{U2}

\item[U3]   \textbf{Connections}: \label{U3}
  \begin{enumerate}
	\item \textbf{Dataflow without connections} (e.g.
\textit{Goto/From} blocks): 
In contrast to Simulink, there is no dataflow without a channel connection in  
connected 
and \textit{consistent}\footnote {Inconsistent SDFGs require
unlimited storage or lead to deadlocks during 
execution\cite{lee_synchronous_1987}.}
SDFGs considered in this paper.

	\item \textbf{Grouping of connections} (e.g. \textit{BusCreator} block 
for 
\textit{bus} signals): 
In Simulink, connections with different properties (e.g. different data types) can 
be grouped into one connection. This is not possible in an SDFG since the tokens 
transfered among a
channel must have the same properties.

      \item \textbf{Connection style}: While in Simulink the storage of data 
between
blocks has the same behavior as that of a register where data can be
overwritten (in case of multi-rate models), the inter-actor communication via 
channels  in
SDFGs follows a (data-flow) FIFO buffer fashion, where tokens must be first 
consumed
before being able to buffer new ones.
\end{enumerate}

\item[U4] \textbf{Sampling rates}: In addition to the number of data transported 
over a
connection by every block activation, a periodic sampling rate is assigned to each 
block
in Simulink to mark its periodic activation at this specific frequency. If all
blocks exhibit the same sampling periods in a model, then this model is called a
\textit{single-rate} model otherwise it is a \textit{multi-rate} model. In SDFGs,
however, an actor is only activated based on the availability of inputs. Actors do 
not
have explicit sampling periods and therefore data rates can only be represented by 
the
rates assigned to their (input/output) ports. \label{Multi-rate}
\end{description}

Because of the above differences, some constraints must be imposed on the
Simulink input model in order to enable its translation to an equivalent SDFG, 
which we
will discuss in the following section.

\subsection{Constraints on the Simulink Model}
\label{subsec:constraints}
 Only Simulink models with fixed-step solver are supported in the translation. In 
case of multi-rates, rate transitions should be inserted to the Simulink model and 
the rates
should be harmonic (divisible). These constraints are indispensable to enable deterministic
code generation
\cite{Simulink-coder-guidlines, bostrom_contract-based_2015}, since we aim with the help of 
Simulink built-in
code-generator to generate SDF-compatible executable code for the translated SDF 
application.  
Even though it is possible to translate a Simulink model to multiple SDFGs, we deal only 
with one application (implemented in
Simulink) at a time in this paper, which results after translation into one 
equivalent SDFG. 
This application is considered to be a control application having the general
structure depicted in Fig.~\ref{fig:codegeneration}.
Moreover, a correct functional simulation of the Simulink model is a 
prerequisite for the
translation in order to get an executable SDFG. 
In addition to above general prerequisites, the following constraints are imposed 
on the
input Simulink model to enable the translation:

\begin{description}
\item[E1] \textbf{Hierarchy:} 
\noindent
Hierarchical blocks  (e.g. \textit{subsystems}), in which one or more functional 
blocks of 
the
types described in \textit{U3-1} and \textit{U3-2} exist,  are not allowed to be
translated to atomic actors. Either these blocks should be removed from the entry 
Simulink model (for they serve only visualization improvement purpose) or the 
model 
should be dissolved at the hierarchy level at which these components exist 
where these blocks are translated and connected in
accordance with the rest of the SDFG.  This constraint is
mandatory, otherwise if we allow an atomic translation of such hierarchical 
functional
blocks, their contained functional blocks of the form \textit{U3-1} and 
\textit{U3-2},
which may be connected with functional blocks in different hierarchical levels, 
would
disappear in the target SDFG. A translation of these blocks would thus no longer 
be
possible and would cause a malfunction of the target SDFG (see restriction 
\textit{E3}).

\item[E2] \textbf{Control-flow logic/Conditional:}
Blocks such as \textit{Triggered/Enabled} subsystems can be translated just like 
the
general subsystems. Upon dissolving the hierarchy of such subsystems, the control
flow takes place now within the atomic functionality of the actor without being in
contradiction to SDFG semantics (c.f. Sec.~\ref{sec:bg:sdfgs}).
In such a translation, however, additional control channels must be defined (see
Sec.~\ref{subsec:translationsteps}).
Yet, the case described in \textit{U2} must still be prohibited. In order to do 
that, 
there is an option ``allowing different data input sizes'' in Simulink for such 
blocks,
which when disabled, prohibits outputs of variable sizes of a control
block\footnote{According to \cite{mat1} blocks having this option are:
\textit{ActionPort, Stateflow,
Enable/Trigger Subsysteme, Switch, Multiport Switch} and \textit{Manual Switch}.}.
A special case of these blocks is the powerful stateflow supported by Simulink.
In our translation we do not flatten the stateflow block and we always translate 
it into one atomic actor.

\item[E3]  \textbf{Connections} 
\begin{enumerate}
 \item   \textbf{Dataflow without connections:}
For blocks having the same behavior described in \textit{U3-1} (such as From/Goto 
or
DataStoreRead,/DataStoreWrite blocks), we assume that the
source
block (e.g. DataStoreWrite block), intermediate block (e.g. DataStoreMemory block) 
and the
target block (e.g. DataStoreRead block)  which communicate without connections are
available in the input Simulink model.  This constraint is important as Simulink 
allows
instantiating a source blocks without for instantiating for e.g. the sink block.

\item  \textbf{Grouping of connections:} 
In order to  support the translation of Simulink models with blocks having the 
same
behavior as those described in  \textit{U3-2}\footnote{e.g.
\textit{BusCreator/BusSelector},  \textit{Bus Assignment} and \textit{Merge} 
blocks
\cite{mat1}.}, two constraints must be imposed.  The first one is that every block
which groups multiple signals (e.g. BusCreator) into one signal must be directly 
connected
to a block which have the opposite functionality (e.g. BusSelector). The second 
constraint
is
imposed on the block (e.g. BusSelector) which takes the grouped signals and splits 
them
again. An ``Output as bus'' should be prohibited in the options of this block.
By doing this, grouping of signals for better visibility in the Simulink model is 
still 
with the limitation above allowed, while prohibiting grouping of signals of 
different 
parameters in one signal in the target translation.

\end{enumerate}
\end{description}

\begin{figure}[h]
\centering 
\includegraphics[width=0.5\textwidth]{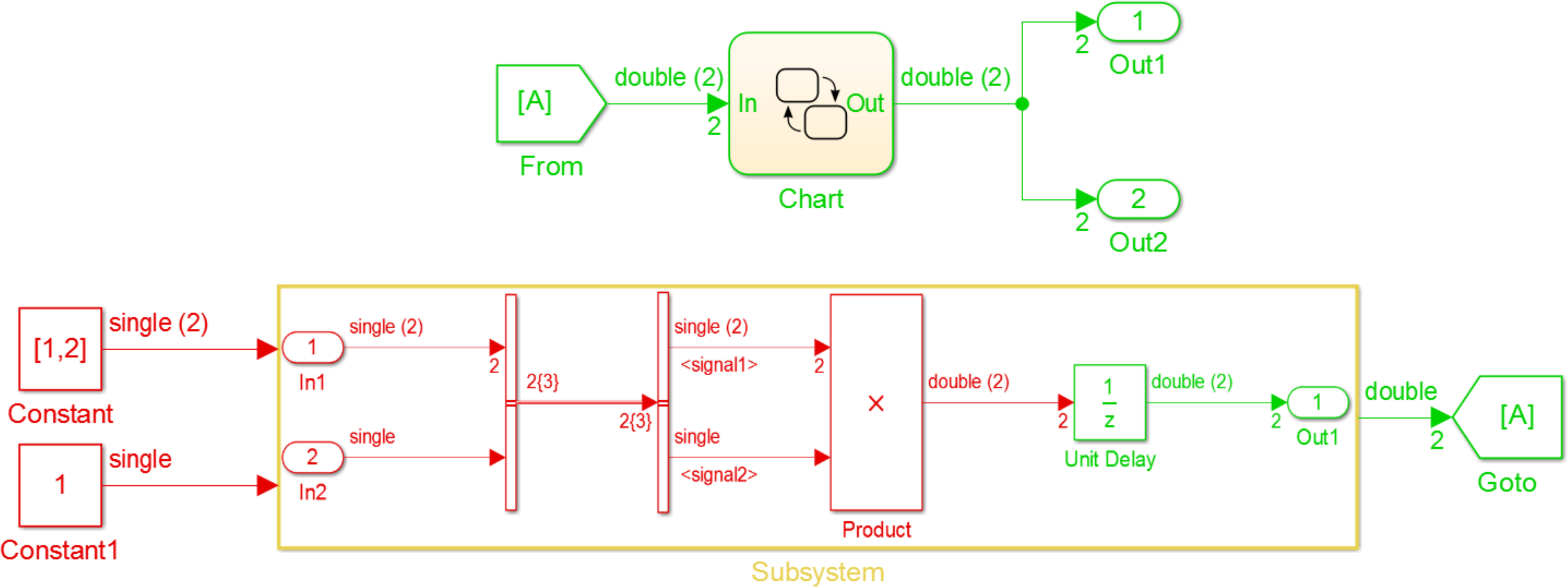}
\caption{Original Simulink model: \textit{Red} having a sample time of 2, \textit{Green} 
having a sample time 4}
\label{fig:transOriginal}
\end{figure}

\subsection{Translation Procedure}
\label{subsec:translationsteps}
\begin{figure*}[htb]
	\begin{center}
		\includegraphics[width=0.75\textwidth]{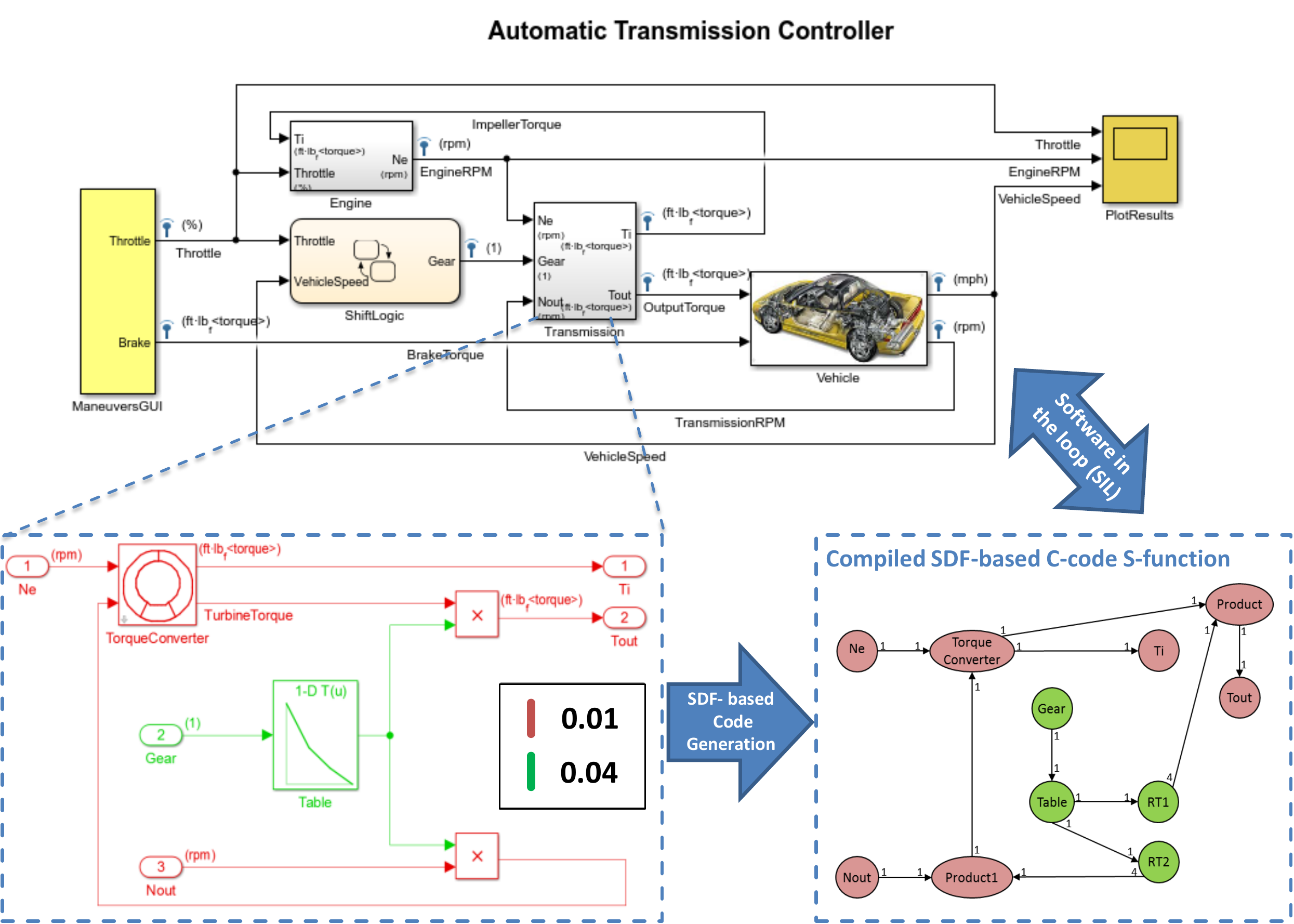}
	\end{center}
	\caption[Motorcontrol]{SDF code-generation of the transmission controller 
model \cite{SIL_Transmision} (with \textit{slow-to-fast} transitions)}
	\label{fig:TransSim2SDF}
\end{figure*}

In the following, we will roughly describe the translation procedure implemented to 
extract an SDFG from a Simulink model with the help of an academical Simulink 
multirate example in Fig.~\ref{fig:transOriginal}. For the translation two main 
phases are required: the \textit{pre-translation} phase where the original Simulink model 
is prepared and checked for the above defined constraints and the \textit{translation} 
phase where the  translation takes place.
\begin{enumerate}
 \item \textbf{Pre-Translation phase:}
\begin{enumerate}
\item \textbf{Checking Requirements:}
Here the Simulink model is checked if it fulfills the 
constraints described above. If this is not the case the translation is aborted with an 
output of the list of unfulfilled constraints.

\begin{figure}[h]
\centering 
\includegraphics[width=0.5\textwidth]{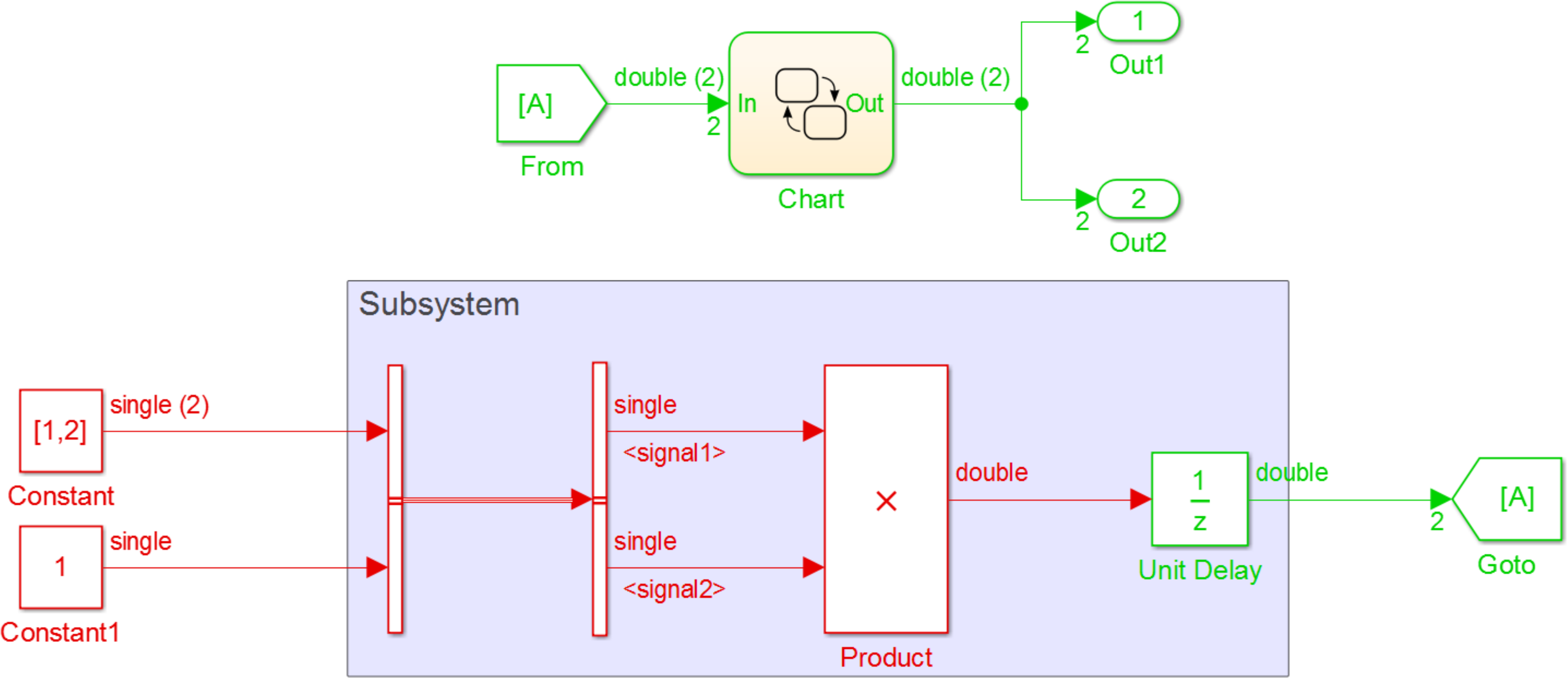}
\caption{Dissolving hierarchy to the desired level}
\label{fig:transhierachy}
\end{figure}

	\item \textbf{Dissolving hierarchy:}
	In this step, a top-down flattening of the Simulink model (respecting
\textit{E1}), till the required
depth level is reached, is done (see Fig.~\ref{fig:transhierachy}).

	\item \textbf{Removing connecting blocks of type U3-1/U3-2:}
Here, blocks respecting the \textit{E3-1/E3-2}  constraint 
are removed. When doing this, the predecessor block of the source block (e.g. 
DataStoreWrite block) is directly connected either to the intermediate (if existent) block 
 (e.g. DataMemory block) or to the successor block of the target block  (e.g. 
DataStoreRead block) and these connecting blocks (source and target blocks) are removed 
(see 
Fig.~\ref{fig:transBus} where BusCreator/BusSelector and Goto/From blocks are removed).


\begin{figure}[h]
\centering 
\includegraphics[width=0.5\textwidth]{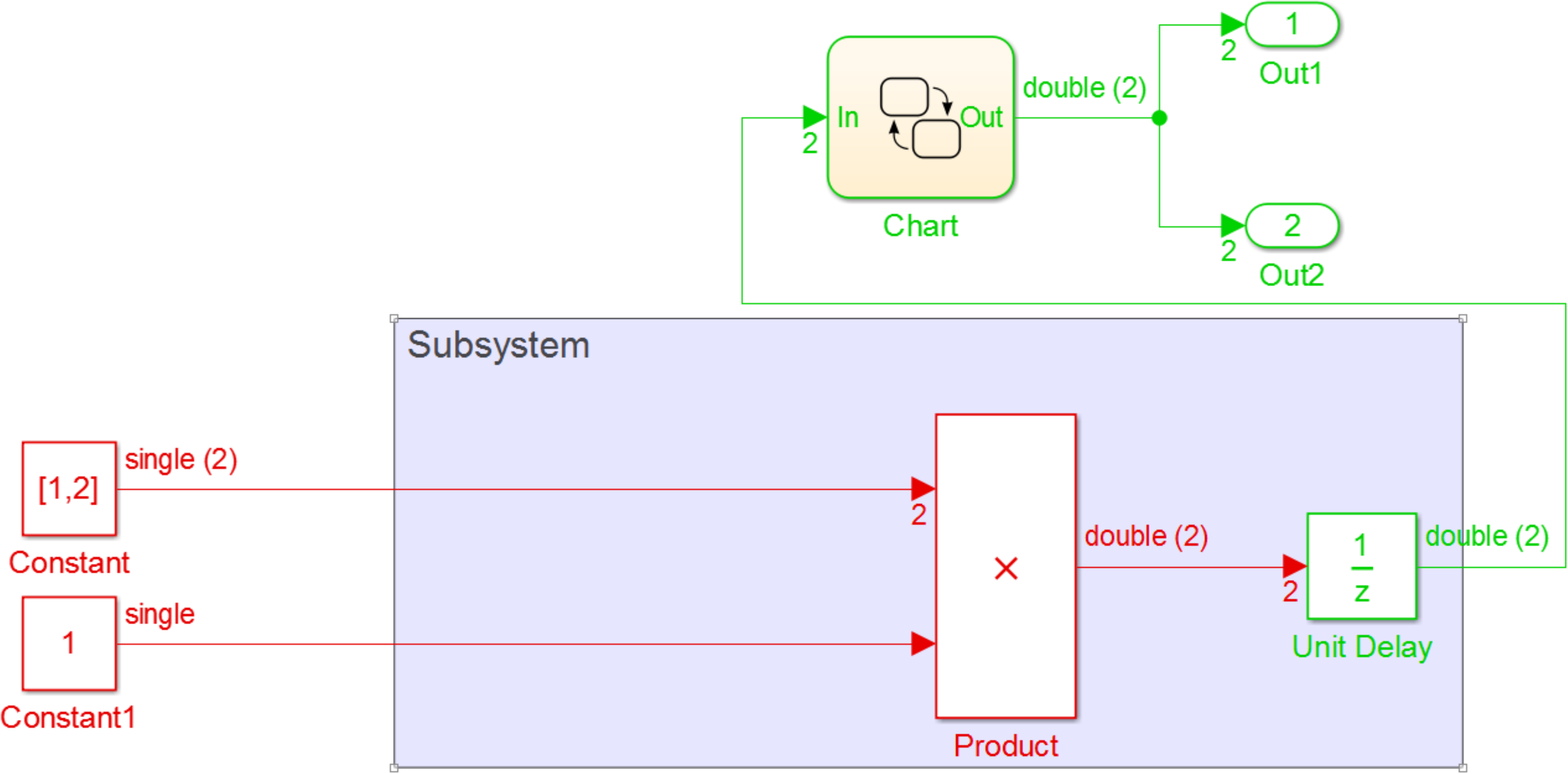}
\caption{Removing connecting blocks of type \textit{U3-1/U3-2}}
\label{fig:transBus}
\end{figure}

\item \textbf{Inserting rate-transition blocks:}
Here rate-transition block are inserted between blocks connected to each other 
and having different sample rates (see Fig.~\ref{fig:transRateTrans}).

\begin{figure}[h]
\centering 
\includegraphics[width=0.5\textwidth]{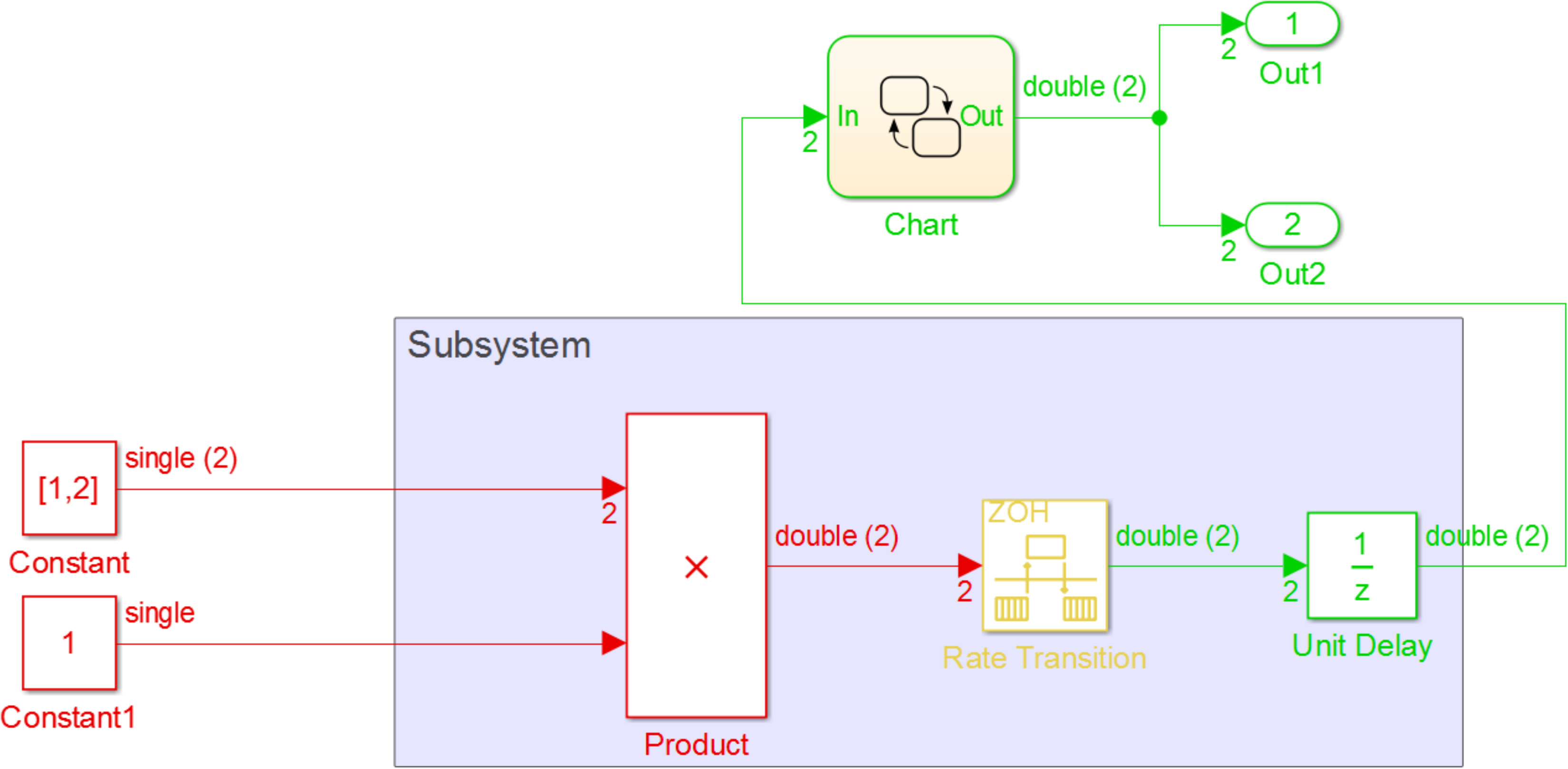}
\caption{Inserting rate-transition blocks between blocks of different sample rates}
\label{fig:transRateTrans}
\end{figure}
\end{enumerate}

\item \textbf{Translation phase:}
In this step, the modified Simulink model is directly translated into an SDFG (see 
Fig.~\ref{fig:sdfg_representation}) according to the following procedure:
\begin{enumerate}
\item \textbf{Translation of blocks:}
If $B$ is the set of all blocks in Simulink model $M$ 
then each block $b_l \in B$ 
in $M$ is
translated into an unique actor in the translated SDFG $a_l\in 
\mathcal{A}$ (where $\mathcal{A}$ is the set of actors see 
Fig.~\ref{fig:sdfg_representation}).

\item \textbf{Translation of connections:}
	Each output port $b_l.o$ is translated into a unique output port 
$a_l.p_o$ and
each input port $b_l.i$ is translated into a unique input port $a_l.p_i$.  
In case multiple connections $t_1, t_2, \cdots, t_n$ going out from an output port
$p_{o1}$ in Simulink (which is permitted in Simulink but not in the SDFG, see connections 
of \textit{statechart} before 
Fig.~\ref{fig:transOriginal} and after translation Fig.~\ref{fig:sdfg_representation}), 
then for each one
of these connections, the output port is replicated $p_{o11}, p_{o12}, \cdots,  
p_{o1m}$
(each having the same properties) in the resulting SDFG, in order to guarantee 
that every channel
$d\in \mathcal{D}$ (set of all channels in an SDFG) has unique input and output ports.
	Now, each  connection $t \in M$ in the Simulink model is translated into 
a channel
$d\in \mathcal{D}$ in the SDFG (see  Fig.~\ref{fig:sdfg_representation}).


	\item \textbf{Extraction of Tokens' sizes and types}: The number of 
the data transfered over a connection represents the size of a token produced/consumed 
when an actor fires (e.g. \texttt{Constant} actor produces a token of size $2$ in 
Fig.~\ref{fig:sdfg_representation}) and their data type represents the data type of that 
token (e.g. \texttt{double} in Fig.~\ref{fig:sdfg_representation}). These parameters can 
be extracted from the model for every connection.

    \item \textbf{Handling Multi-rates}: The following method for handling multirates was inspired
	from \cite{bostrom_contract-based_2015, zhang_bridging_2013}.
	To determine the rates of the actors' input and 
	output ports we must differentiate between three cases: \textit{fast-to-slow} transition, 
\textit{slow-to-fast} transitions and transitions between blocks having the same rates.
For the latter case, source and destination actors are denoted by a rate of \texttt{1} on 
their ports indicating 
the production/consumption of one token (of specific size per channel) whenever 
activated.
In case of \textit{slow-to-fast} transition (see e.g. in Fig.~\ref{fig:Multirate-s2f} and 
in Fig.~\ref{fig:TransSim2SDF}), the rate of the output port of the rate-transition actor 
\begin{equation}
 R.p_o.rate = b_{src}.sp/b_{dst}.sp,
\end{equation}
 where $p_o$ is the output port of the actor $R$, $b_{src}$ and $b_{dst}$ 
are the source and destination blocks connected via rate-transition block and $sp$ the 
sample time of the corresponding block. The rate of the input port of $R$ is set to 
\texttt{1}. This basically realizes multiple copies of tokens of the slower actor for the 
faster actor to run. 
\begin{figure}[t]
\centering 
\includegraphics[width=0.4\textwidth]{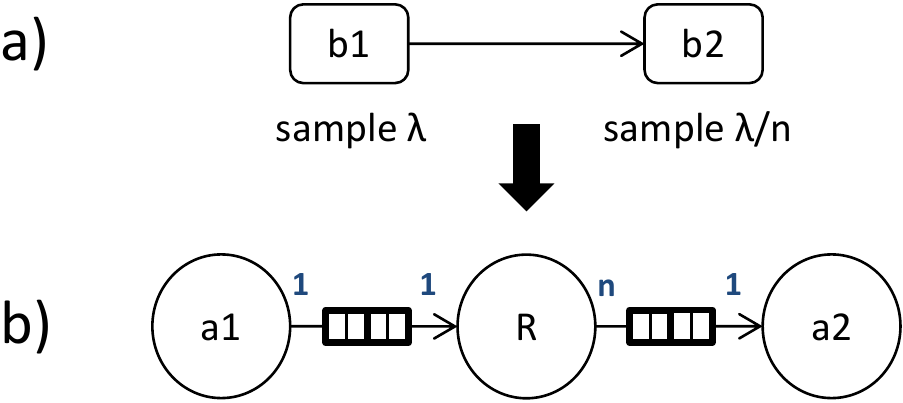}
\caption{Example of a Simulink slow-to-fast multirate model shown in (a). By adding a 
rate-transition actor $R$, a valid translation to SDFG can be achieved in (b).}
\label{fig:Multirate-s2f}
\end{figure}
\begin{figure}[t]
\centering 
\includegraphics[width=0.4\textwidth]{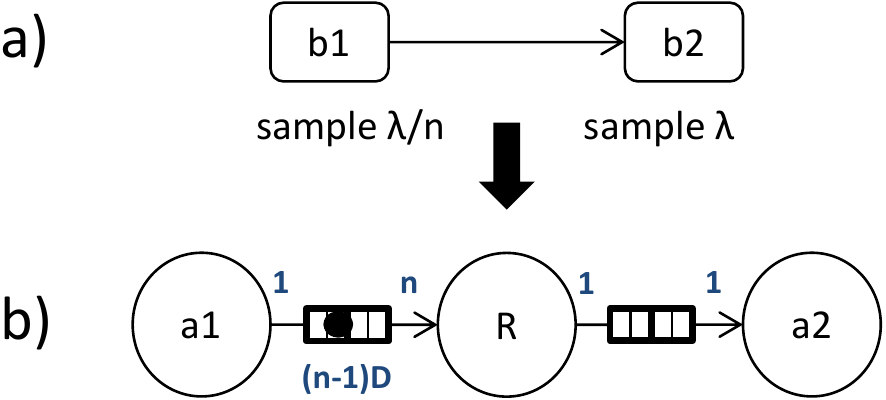}
\caption{Example of a Simulink fast-to-slow multirate model (a) and its equivalent 
SDFG in (b).}
\label{fig:Multirate-f2s}
\end{figure}

\begin{figure}[tb]
\centering 
\includegraphics[width=0.45\textwidth]{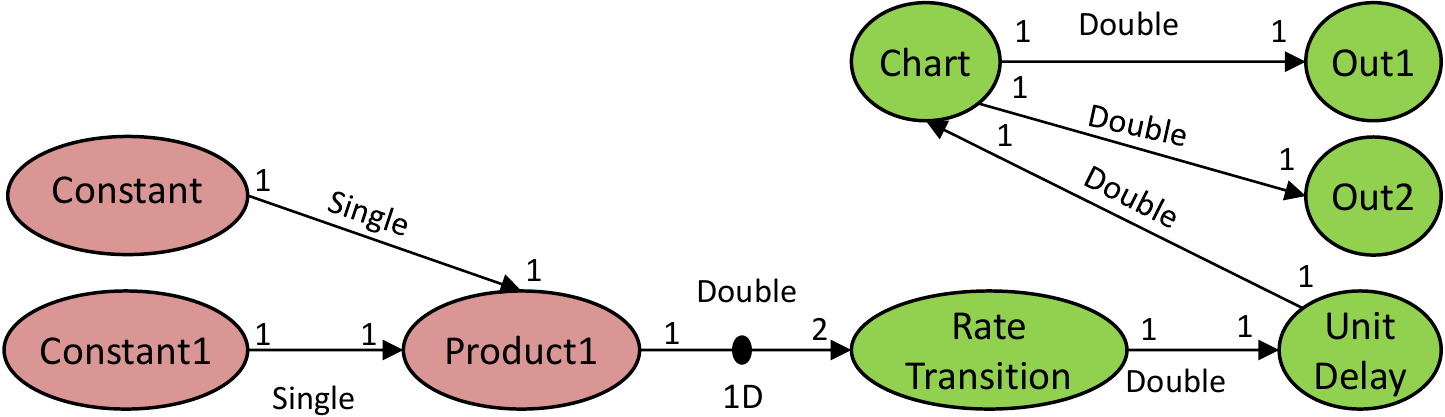}
\caption{Translating modified Simulink model with \textit{fast-to-slow} transitions into 
SDF}
\label{fig:sdfg_representation}
\end{figure}

\begin{figure*}[t]
\centering 
\includegraphics[width=0.8\textwidth]{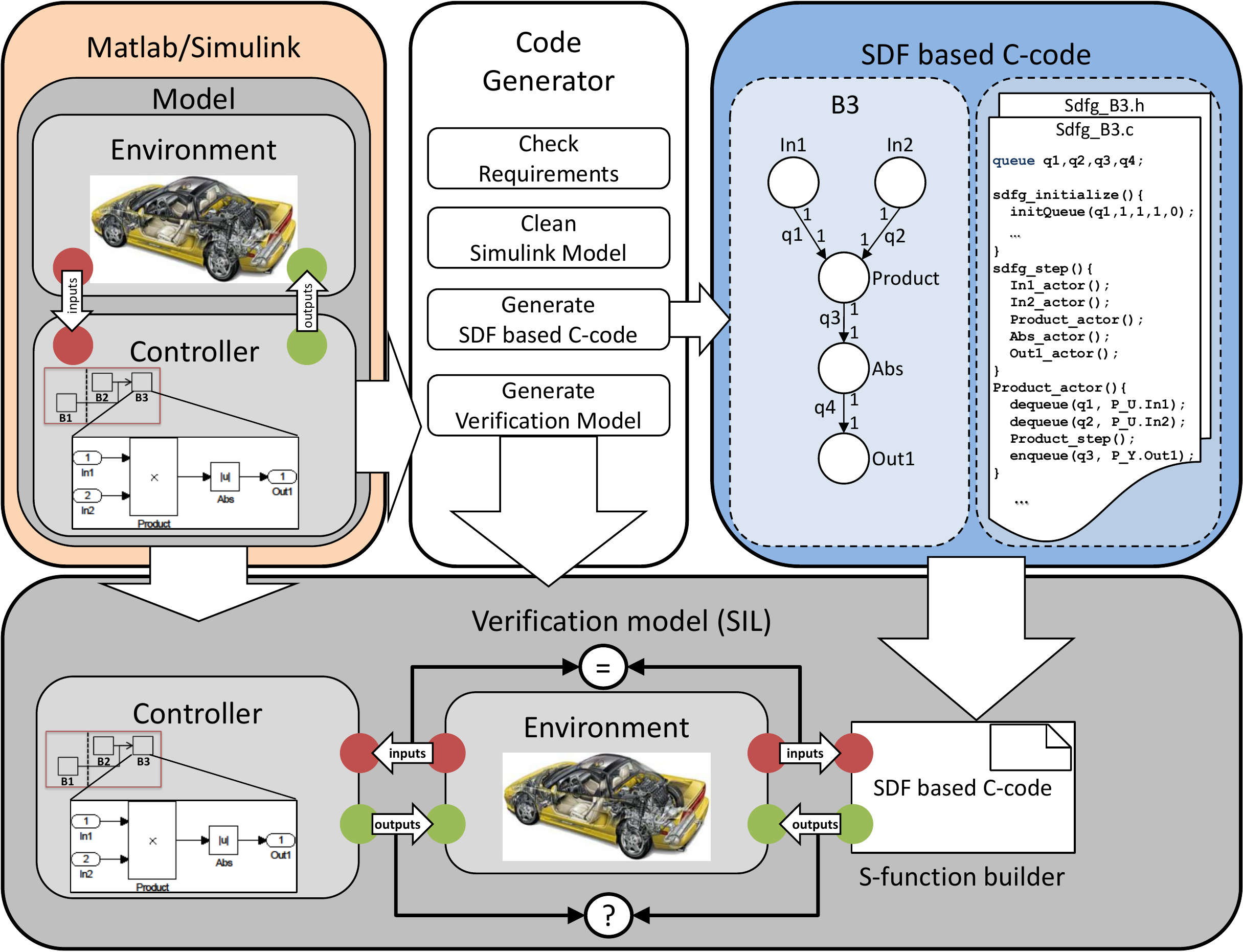}
\caption{Structure of the model transformation and code-generation framework. 
\textit{Simulink Model:} consisting mainly of a controller and the environment model. 
\textit{Code 
generator:} implementing the translation of Simulink models to SDFGs, generating SDF 
code and the verification model. In this figure, the 
transformation is exemplary applied on the sub-block \texttt{B3} of the 
controller Simulink model. \textit{SDF based C-code:} executable code 
of the generated SDFG. \textit{Verification model:} consists of the reference 
controller and 
environment models with an extra \textit{S-function builder} block in which the generated 
SDF code 
is embedded and connected to the environment allowing SIL verification.}
\label{fig:codegeneration}
\end{figure*}

In case of \textit{fast-to-slow} 
transition, the rate $R.p_i.rate$ of the input port ($p_i$) of the rate-transition actor  
$R$ can be calculated as follows: 
\begin{equation}
R.p_i.rate = b_{dst}.sp/b_{src}.sp,
\end{equation}
The output port rate is set to 1. This mainly accumulates tokens on the rate-transition 
actor 
and outputs the most freshest token of the faster actor.
Furthermore, in this case 
a number of delay tokens equal to:
\begin{equation}
d.delay =  (b_{dst}.sp/b_{src}.sp)-1,
\end{equation}
are placed on the input channel $d\in \mathcal{D}$) of the 
rate-transition actor in order to enable considering the initial token produced by the 
fast actors at the first firing (see 
Fig.~\ref{fig:Multirate-f2s} and Fig.~\ref{fig:sdfg_representation}).

\item \textbf{Adding event channels:} if the subsystem is a triggered one then, depending 
on the hierarchy level chosen, extra connections are added in this step for handling 
(enabling/triggering) events. These edges are needed when the
hierarchy of a enabled/triggered subsystem is dissolved. In this case,
each block, belonging to the triggered or enabled subsystem has to be sensitive to the 
(triggering/enabling) event and thus is connected with the event source.
\end{enumerate}
\end{enumerate}

Finally, the actors in the resulting SDF graph can be statically scheduled to 
obtain a minimal periodic static schedule (\texttt{Constant} \texttt{Constant1} 
\texttt{Product} )$^2$ (\texttt{RateTransition} \texttt{UnitDelay} \texttt{Chart} 
\texttt{Out1} \texttt{Out2}).

\subsection{Code-generation and SIL Simulation}
\label{subsec:sil}

After describing the translation procedure of Simulink models into SDFGs, we will 
describe in the following the corresponding implementation on top of Simulink 
and how to utilize Simulink code-generator to enable SDF code generation and SIL 
verification.
Generating an equivalent SDF-compatible C code is useful to verify the functional 
equivalence between Simulink models and the generated SDFGs on one side, and to enable the 
direct code deployment on target hardware platforms, on the other side. 

Fig.~\ref{fig:codegeneration} shows the different steps involved in the model 
transformation process within our code-generation framework. 
The code generator constitutes the major part of our model transformation, taking the 
Simulink model as an input and generating the SDF code and the verification (SIL) model as 
output. We implemented the code generator as a Matlab script taking use of the Matlab 
API to manipulate and extract needed information from Simulink models. The implemented 
code generator constitutes mainly of the following functions:
\begin{itemize}
 \item \textit{Check Requirements:} the Simulink
model is checked if it fulfills the constraints described Sect.\ref{subsec:constraints}. 
For e.g. in case of multirates, the rates are checked whether or not these are 
integers and divisible.
 \item \textit{Clean Model:} in this step, the chosen subsystem (to be translated) is 
restructured according to the pre-translation phase (see 
Sect.~\ref{subsec:translationsteps}): hierarchies dissolved, routing blocks dissolved and 
rate-transition blocks inserted. In addition, every block of the desirable hierarchy is 
packaged in a subsystem and the connections are updated since the code-generation is only 
possible for subsystems.
 \item \textit{Generate SDF Code:} this function uses the  
\textit{Simulink Embedded Coder} and an SDF API to generate SDF-based embedded C code from 
the modified model of the previous step (see example at the right of 
Fig.~\ref{fig:codegeneration}).
In this case, embedded C code is first generated for each block at the the chosen 
hierarchy level. The SDF-based C code is generated by using the predefined SDF library 
files (\texttt{SDFLib.h, SDFLib.c} implemented according to description in 
\cite{SDFImplementation2013}) that have been already loaded into the folder structure. The 
output are two files (\texttt{sdfg\_<Name>.h, sdfg\_<Name>.c}) for every SDFG, in which 
the actors and channels are defined and instantiated according to the translation concept.
For each actor a corresponding function is generated (E.g. \texttt{Product\_actor()} see 
Fig.~\ref{fig:codegeneration}), in which 
data availability of every 
input channel is checked (implemented as FIFO \texttt{queue}) and, if all inputs are read 
(E.g. \texttt{dequeue(q1, P\_U.In1)}), the 
actor executes its internal computation behavior (implemented in a step function for e.g. 
\texttt{Product\_step()}) and the results are written into 
its output channels (E.g. \texttt{enqueue(q3, P\_Y.Out1)}). 
In addition, a basic valid static schedule is generated and implemented for the SDFG 
(see \texttt{sdfg\_step()} in Fig.~\ref{fig:codegeneration}).

 \item \textit{Generate Verification model:}
the latest step targets the realization of a SIL simulation (see bottom-right of 
Fig.~\ref{fig:codegeneration}). For this, we further enhance the code generator to allow 
the automatic integration of the generated SDF-compatible code into a C file of an 
\textit{S-function} block. The S-function block is then automatically generated and 
inserted into a new-created verification model. The verification model includes also the 
original subsystem (controller) with the environment model. The S-function has the same 
interfaces as the original subsystem which allows a seamless SIL simulation with the 
environment model.  Doing this, the functional equivalence of the translated model and the 
original one, can be verified automatically. 
\end{itemize}

   \begin{figure*}[!ht]
     \subfloat[Model-In-the-Loop Simulation Results\label{subfig-1:MIL-Trans}]{%
       \includegraphics[width=0.5\textwidth]{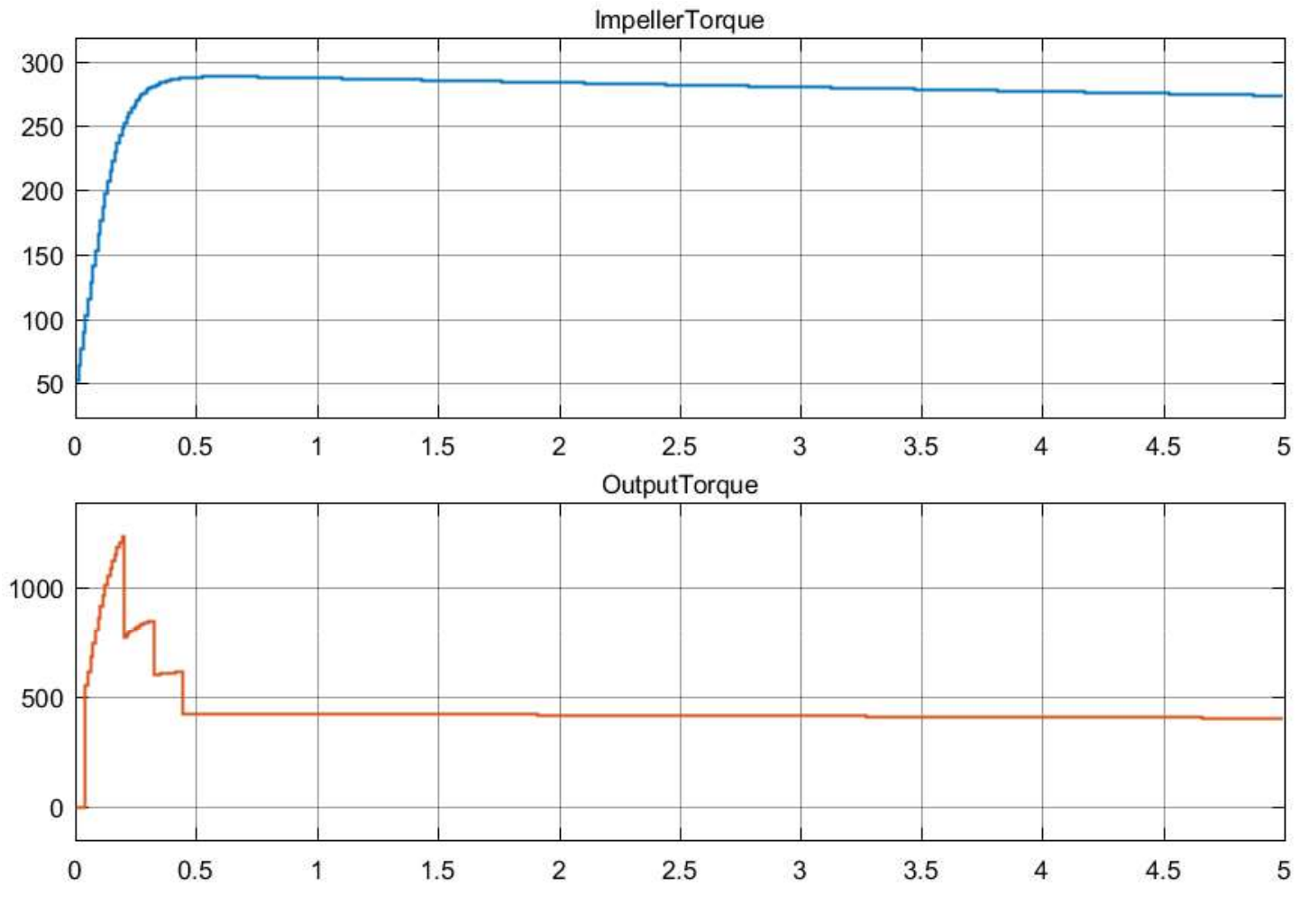}
     }
     \hfill
     \subfloat[Software-In-the-Loop Simulation Results\label{subfig-2:SIL-Trans}]{%
       \includegraphics[width=0.5\textwidth]{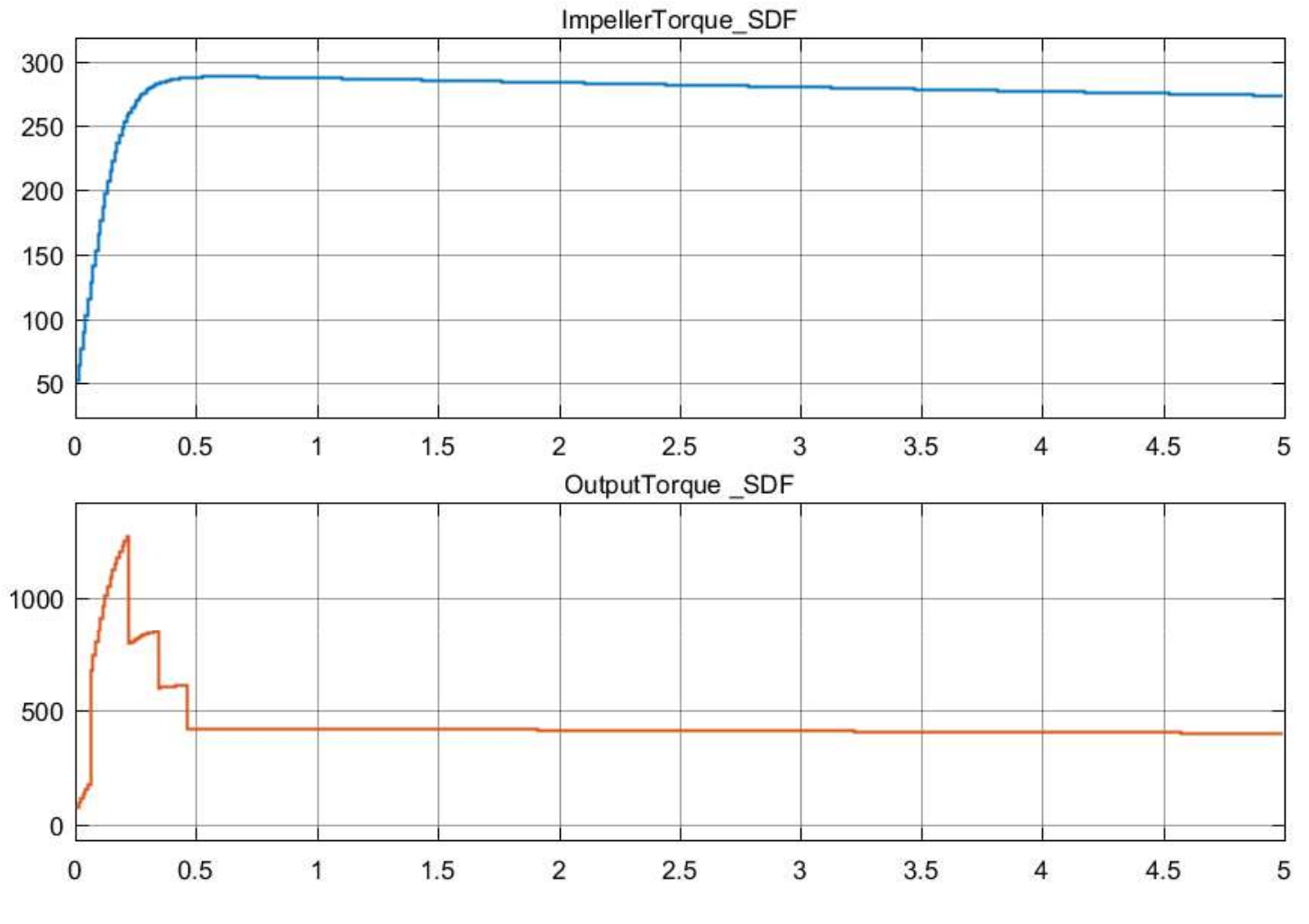}
     }
     \caption{Verification results of the Transmission control model showing equivalent 
outputs of the SIL  (see Fig.~\ref{subfig-2:SIL-Trans}) and the MIL  (see 
Fig.~\ref{subfig-1:MIL-Trans}) simulations.}
     \label{fig:TransmissionResults}
   \end{figure*}

 \begin{figure*}[htb]
	\begin{center}
		\includegraphics[width=0.97\textwidth]{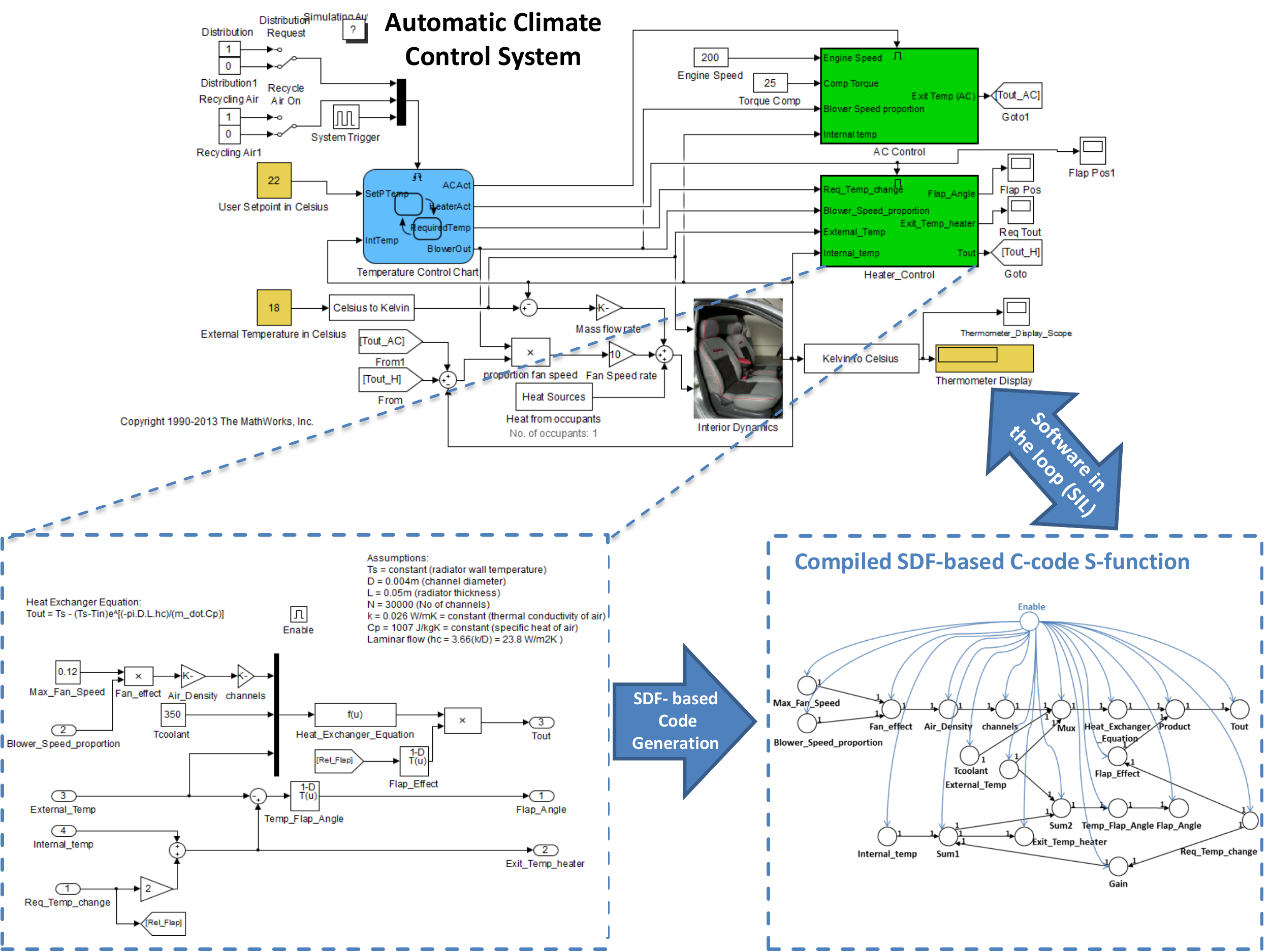}
	\end{center}
	\caption[Motorcontrol]{SDF code-generation of the single-rate \textit{triggered} 
Climate controller model \cite{SIL_Heat}}
	\label{fig:HeatSim2SDF}
\end{figure*}
 
   \begin{figure*}[!ht]
     \subfloat[Model-In-the-Loop Simulation Results\label{subfig-1:MIL-Heat}]{%
       \includegraphics[width=0.5\textwidth]{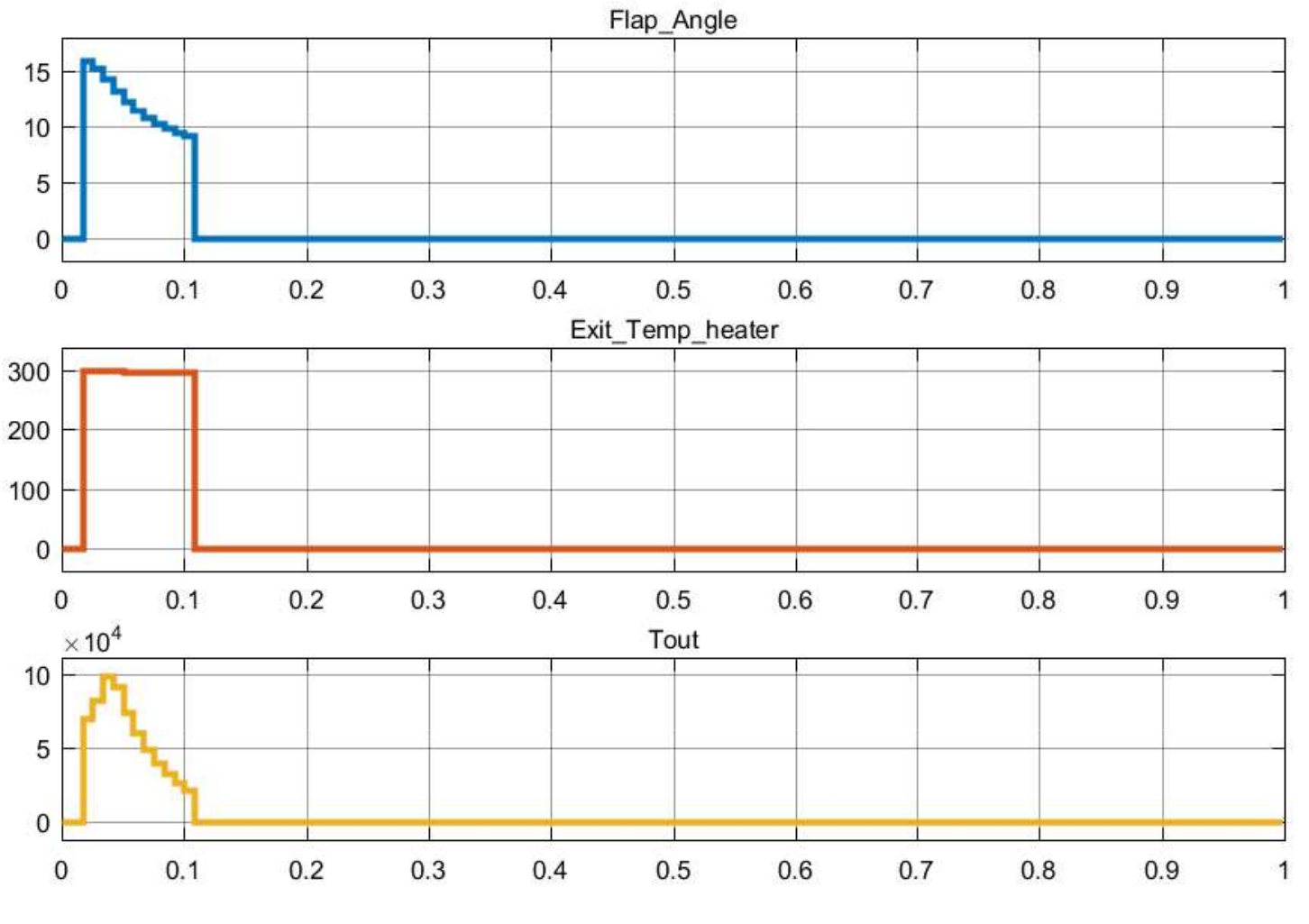}
     }
     \hfill
     \subfloat[Software-In-the-Loop Simulation Results\label{subfig-2:SIL-Heat}]{%
       \includegraphics[width=0.5\textwidth]{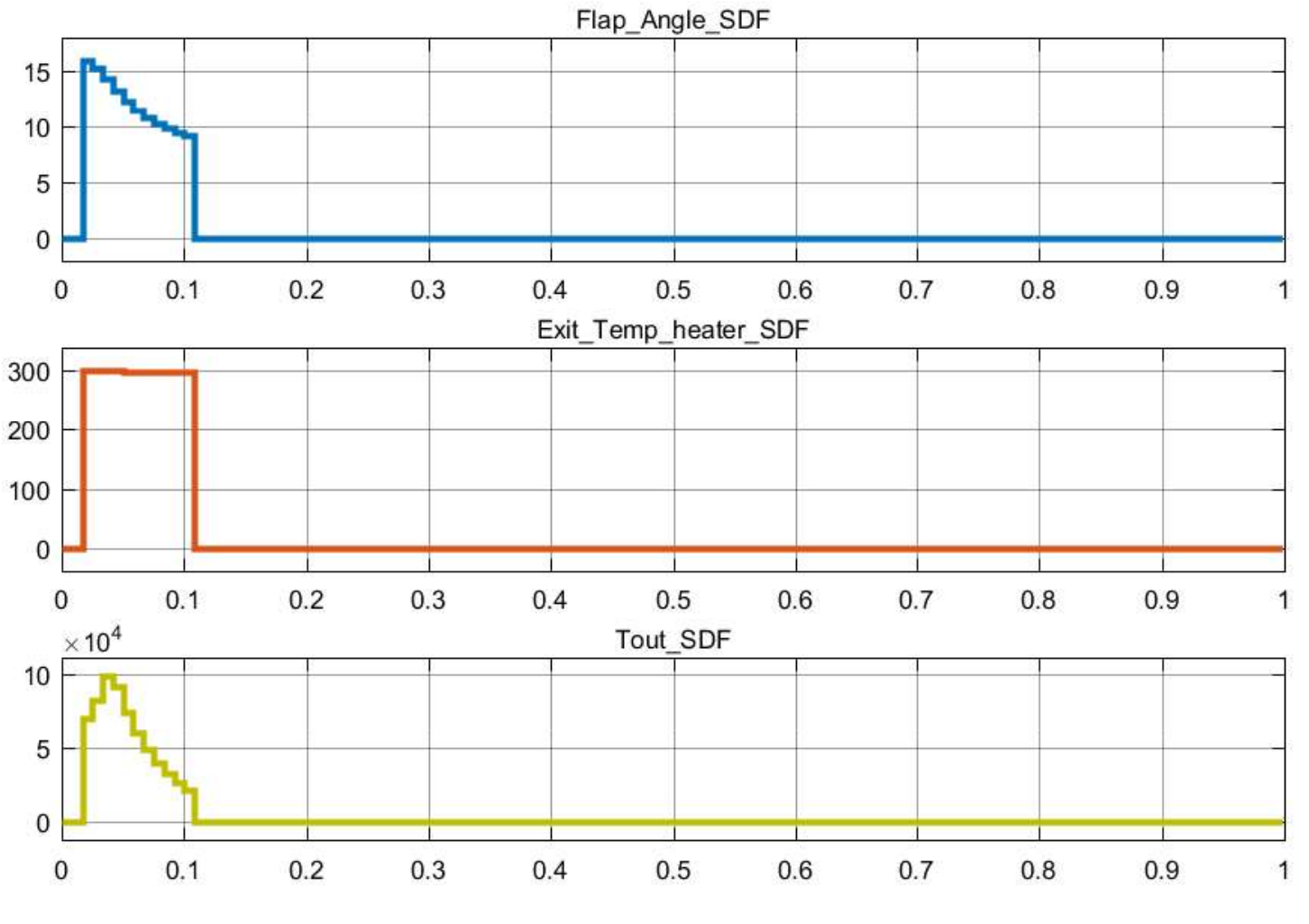}
     }
     \caption{Verification results of the Heat control model showing equivalent outputs of 
the SIL 
(see 
Fig.~\ref{subfig-2:SIL-Heat}) and  the MIL (see 
Fig.~\ref{subfig-1:MIL-Heat}) simulations.}
     \label{fig:HeatResults}
   \end{figure*}

\section{Evaluation}\label{Evaluation}
We have conducted two experiments to demonstrate the viability of our approach 
being able of translating a Transmission Controller Unit (TCU) model (c.f. 
\cite{SIL_Transmision}) and a Climate Controller model (c.f. 
\cite{SIL_Heat}) each to a corresponding SDFG and to 
generate for each case an equivalent SDF C code.

The TCU model depicted in Fig.~\ref{fig:TransSim2SDF} is a typical model exhibiting 
multirates.
The translation for the TCU subsystem (seen at the bottom of Fig.~\ref{fig:TransSim2SDF}) 
was straightforward since the model respected (per construction) the constraints made in 
Sect.~\ref{subsec:translationsteps}.
Fig.~\ref{fig:TransmissionResults} shows that the outputs (impeller torque, output torque) 
of both the reference TCU and the generated SDF-compatible TCU code are equivalent.

More complexity is exhibited by the Automatic Climate Control System (seen 
in Fig.~\ref{fig:HeatSim2SDF}), where the \textit{Heater controller} subsystem was 
translated. In 
addition to the variety of blocks used, the Heater subsystem is a triggered subsystem which 
only executes when the enable signal is true. 
As seen in the generated SDFG (see Fig.~\ref{fig:HeatSim2SDF}), the \texttt{Enable} actor 
is connected via extra-created channels to all actors within the \textit{Heater\_Control} 
SDFG. Only if a true value arrives at these dedicated channels, 
then the corresponding actor will be activated to perform its internal computation. If 
this is not the case, the actor will read its input queues, skip the computation part 
(step function) and update output queues with values of the previous step results.
Also the SIL and MIL results of this experiment show equivalent values as depicted in 
Fig.~\ref{fig:HeatResults} concluding a functionally equivalent SDF code-generation.

\section{Conclusion}
In this work, a translation approach for Simulink models (respecting defined rules) 
to SDFGs was presented. 
Thanks to the automated code-generation of SDF code from the original Simulink 
model and 
the Software-in-the-loop simulation, tests can be automated to show the functional 
equivalence of this translation.
The translation was demonstrated successfully with a medium-sized Transmission Controller 
Unit model from the automotive domain and with a Climate Controller use-case.
In future work, we will take a look at the possibility of optimizing the 
code-generation of Simulink models for MPSoCs. For this, we can take use of the generated 
SDF code and mature optimizing/parallelization techniques from the SDF 
research domain \cite{bhattacharyya_clustering, bhattacharyya_synthesis_1999} to 
enable efficient implementations of embedded systems.

 \section{Acknowledgments}
   This work has been partially supported by the SAFEPOWER project with funding 
 from the European Union's Horizon 2020 research and innovation programme under 
 grant agreement No 646531.
 
 \newpage
 
\bibliographystyle{abbrv}
\bibliography{literatur}

\begin{thebibliography}{10}

\bibitem{bhattacharyya_clustering}
S.~Bhattacharyya, P.~Murthy, and E.~Lee.
\newblock {APGAN} and {RPMC}: Complementary heuristics for translating {DSP}
  block diagrams into efficient software implementations.
\newblock {\em Design Automation for Embedded Systems}, 2(1):33--60, 1997.

\bibitem{bhattacharyya_synthesis_1999}
S.~Bhattacharyya, P.~Murthy, and E.~Lee.
\newblock Synthesis of embedded software from synchronous dataflow
  specifications.
\newblock {\em The Journal of VLSI Signal Processing}, 21(2):151--166, 1999.

\bibitem{bostrom_contract-based_2015}
P.~Bost\"orm and J.~Wiik.
\newblock Contract-based verification of discrete-time multi-rate {Simulink}
  models.
\newblock {\em Software \& Systems Modeling}, pages 1--21, June 2015.

\bibitem{buker_automated_2013}
M.~B\"uker.
\newblock {\em An {Automated} {Semantic}-{Based} {Approach} for {Creating}
  {Task} {Structures}}.
\newblock Dissertation, University of Oldenburg, 2013.

\bibitem{caspi_Simulink_2003}
P.~Caspi, A.~Curic, A.~Maignan, C.~Sofronis, S.~Tripakis, and P.~Niebert.
\newblock From {Simulink} to {SCADE}/{Lustre} to {TTA}: a layered approach for
  distributed embedded applications.
\newblock In {\em {ACM} {Sigplan} {Notices}}, volume~38, pages 153--162, 2003.

\bibitem{Simulink2sdfBsc_2011}
C.~Dominik.
\newblock Conception and {Implementation} of {Parallelism} {Analyses} in
  {MATLAB}/{SIMULINK} {Models} for programming {Embedded}
  {Multicore}-{Systems}.
\newblock Bsc. thesis, TU M\"unchen, 2011.

\bibitem{fakihJSA2015}
M.~Fakih, K.~Gr{\"{u}}ttner, M.~Fr{\"{a}}nzle, and A.~Rettberg.
\newblock State-based real-time analysis of {SDF} applications on {MPSoCs} with
  shared communication resources.
\newblock {\em Journal of Systems Architecture - Embedded Systems Design},
  61(9):486--509, 2015.

\bibitem{gajski}
D.~D. Gajski, S.~Abdi, A.~Gerstlauer, and G.~Schirner.
\newblock {\em Embedded System Design: Modeling, Synthesis and Verification}.
\newblock Springer Science \& Business Media, aug 2009.

\bibitem{rtasKlikpoKM16}
E.~C. Klikpo, J.~Khatib, and A.~M. Kordon.
\newblock Modeling multi-periodic simulink systems by synchronous dataflow
  graphs.
\newblock In {\em 2016 {IEEE} Real-Time and Embedded Technology and
  Applications Symposium (RTAS), Vienna, Austria, April 11-14, 2016}, pages
  209--218, 2016.

\bibitem{lee_synchronous_1987}
E.~Lee and D.~Messerschmitt.
\newblock Synchronous data flow.
\newblock {\em Proceedings of the IEEE}, 75(9):1235--1245, 1987.

\bibitem{Ptolemy2011}
E.~A. Lee.
\newblock Heterogeneous actor modeling.
\newblock In {\em Proceedings of the Ninth ACM International Conference on
  Embedded Software}, EMSOFT '11, pages 3--12, New York, NY, USA, 2011. ACM.

\bibitem{LeeM87}
E.~A. Lee and D.~G. Messerschmitt.
\newblock Static scheduling of synchronous data flow programs for digital
  signal processing.
\newblock {\em {IEEE} Trans. Computers}, 36(1):24--35, 1987.

\bibitem{Sim2Modal2Ptolmy2SDF2011}
G.~Li, R.~Zhou, R.~Li, W.~He, G.~Lv, and T.~J. Koo.
\newblock {A Case Study on SDF-Based Code Generation for ECU Software
  Development}.
\newblock In {\em Proceedings of the 2011 IEEE 35th Annual Computer Software
  and Applications Conference Workshops}, COMPSACW'11, pages 211--217, 2011.

\bibitem{GitHubSimulink2SDF}
S.~Li.
\newblock Simulink2sdf - converter.
\newblock \url{https://github.com/zcold/Simulink2SDF}, 2013.
\newblock (GitHub, open-source Code, last accessed on 01.09.2016).

\bibitem{Simulinksdf2}
R.~Lublinerman and S.~Tripakis.
\newblock Translating data flow to synchronous block diagrams.
\newblock In P.~Eles and A.~D. Pimentel, editors, {\em ESTImedia}, pages
  101--106. IEEE, 2008.

\bibitem{SIL_Transmision}
MathWorks.
\newblock Modeling an automatic transmission controller.
\newblock \url{
  https://de.mathworks.com/help/simulink/examples/modeling-an-automatic-transmission-controll
  er.html}.
\newblock (last accessed on 01.09.2016).

\bibitem{SIL_Heat}
MathWorks.
\newblock Simulating automatic climate control systems.
\newblock \url{
  https://de.mathworks.com/help/simulink/examples/simulating-automatic-climate-control-system
  s.html}.
\newblock (last accessed on 01.09.2016).

\bibitem{mat1}
MathWorks.
\newblock Simulink - simulation and model-based design.
\newblock \url{http://de.mathworks.com/products/simulink/blocklist.html}.
\newblock (last accessed on 01.09.2016).

\bibitem{Simulink-coder}
{MathWorks, Inc.}
\newblock {Automatic Code Generation - Simulink Coder}.
\newblock \url{http://www.mathworks.de/products/simulink-coder/}, 2015.
\newblock (last accessed on 01.11.2015).

\bibitem{Simulink-coder-guidlines}
{MathWorks, Inc.}
\newblock {Modeling Guidelines for Code Generation}.
\newblock Technical Report Version 1.10 (Release 2015b), 2015.

\bibitem{stateflow}
{MathWorks, Inc.}
\newblock {Stateflow official website}.
\newblock \url{http://www.mathworks.com/products/stateflow/}, 2015.
\newblock (last accessed on 01.09.2016).

\bibitem{miller_formal_2005}
S.~Miller, E.~Anderson, L.~Wagner, M.~Whalen, and M.~P.~E. Heimdahl.
\newblock Formal verification of flight critical software.
\newblock In {\em Proceedings of the {AIAA} {Guidance}, {Navigation} and
  {Control} {Conference} and {Exhibit}}, 2005.

\bibitem{SDFImplementation2013}
P.~R. Schaumont.
\newblock {\em A Practical Introduction to Hardware/Software Codesign}.
\newblock Springer US, 2013.

\bibitem{simulink2sdfWarsitz2016}
S.~Warsitz and M.~Fakih.
\newblock Simulink-modell-{\"{u}}bersetzung in synchrone datenflussgraphen.
\newblock In {\em 19th {GI/ITG/GMM} Workshop Methoden und Beschreibungssprachen
  zur Modellierung und Verifikation von Schaltungen und Systemen, {MBMV} 2016,
  Freiburg im Breisgau, Germany, March 1-2, 2016.}, pages 89--101, 2016.

\bibitem{zhang_bridging_2013}
L.~Zhang, M.~Glab, N.~Ballmann, and J.~Teich.
\newblock Bridging algorithm and {ESL} design: {Matlab}/{Simulink} model
  transformation and validation.
\newblock In {\em Specification \& {Design} {Languages} ({FDL}), 2013 {Forum}
  on}, pages 1--8, 2013.

\end{thebibliography}

\end{document}